\documentclass[twocolumn,aps,prl,groupedaddress]{revtex4}
\usepackage{amsmath,amssymb}
\usepackage{graphicx}
\usepackage{dcolumn}
\usepackage{bm}
\usepackage{xcolor}
\usepackage{url}
\usepackage{hyperref}
\hypersetup{
    bookmarksnumbered,
    linktocpage=true,
    colorlinks=true,
    bookmarks=true,
    citecolor=blue,
    urlcolor=blue,
    linkcolor=blue,
    citebordercolor=blue,
    urlbordercolor=blue,
    linkbordercolor=blue,
    breaklinks=true,
    pdfpagelabels=true,
    }

\begin{document}

\title{Enhancing the intense field control of molecular fragmentation}

\author{Fatima Anis and B. D. Esry}
\affiliation{J.R. Macdonald Laboratory, Kansas State University, Manhattan, Kansas 66506}
\begin{abstract}
We describe a pump-probe
scheme with which the spatial asymmetry of dissociating molecular fragments 
--- as controlled by the carrier-envelope phase of an intense few-cycle laser pulse ---
can be enhanced by an order of magnitude or more. We illustrate the scheme using extensive, 
full-dimensional calculations for dissociation of H$_2^+$ and include the averaging necessary 
for comparison with experiment.
\end{abstract}
\maketitle

In recent years, considerable experimental effort has been invested in developing the
ability to control chemical reactions with intense, few-cycle laser pulses \cite{Paulus:Nature:2001:CEPATI,Roudnev:PRL:2004:CEP,Kling:Science:2006:ElectronLocalization}. 
The canonical reaction chosen has been molecular dissociation, and a common
measure of the degree of its coherent control  
is the spatial asymmetry of fragments relative to a linearly polarized laser
field. Quantum mechanically, this asymmetry arises from the interference of pathways 
that lead to even and odd parity states \cite{PhysRevLett.92.033002,Roudnev:PRL:2007:CEP,Hua:JPB:2009}.
In strong fields, these pathways can involve many photons, and the relative phase between pathways
--- and thus the outcome --- can be controlled by varying laser parameters such as the carrier 
envelope phase (CEP) \cite{Paulus:Nature:2001:CEPATI,Paulus:PRL.89.153001,Roudnev:PRL:2004:CEP,Kling:Science:2006:ElectronLocalization,Roudnev:PRL:2007:CEP,Hua:JPB:2009,Kling.MolPhys.2008,kremer:213003,nakajima:213001,Anis:AtCEP:FTC:2009,Baltuska:Nature:2003:HHG,PhysRevLett.105.223001,PhysRevLett.103.103002} or the relative phase between different colors \cite{Charron.R641,Charron.PRL.71.692,Charron.PRL.75.2815,he:083002,he:213002,RayTwocolorPRL,SinghPRL.104.023001,Calvert.JPB.43.011001.2010}.
In this Letter, we will focus on control via the CEP.

In the dipole approximation, the CEP $\varphi$ for a Gaussian laser pulse ${\cal E}(t)$ is defined 
from \cite{Param}
\begin{align}
{\cal E}(t) = {\cal E}_{0}e^{-t^2/\tau^2}\cos(\omega t+\varphi).
\label{pulse}
\end{align}
Generally, the largest CEP-dependent asymmetries have been observed for ionized 
electrons~\cite{Paulus:Nature:2001:CEPATI}. The asymmetries for the nuclear fragments resulting
from dissociation have, unfortunately, been much smaller \cite{Kling:Science:2006:ElectronLocalization,Kling.MolPhys.2008,kremer:213003}. 
These weak effects --- combined with the ongoing challenge of producing intense, few-cycle, CEP
stabilized pulses --- greatly limit experimentalists' abilities to measure and explore this intriguing 
means of control. One important recent advance is the ability to measure the CEP of each pulse \cite{Nat.Phys.5.357.2009,PhysRevA.83.013412},
alleviating the need for CEP stability during the measurements. 

While CEP-dependent asymmetric break up of H$_2^+$ was predicted a few
years ago \cite{Roudnev:PRL:2004:CEP}, successful measurements have not yet been
made starting directly from this benchmark system, {\em e.g.} in an ion beam 
experiment \cite{BenItzik:PRL:2005:H2}. Experiments have instead begun with the more complicated 
H$_2$ \cite{Kling:Science:2006:ElectronLocalization,kremer:213003}. 
With only one electron, the number of control pathways for H$_2^+$ is smaller than for H$_2$ making 
the interpretation more straightforward. Moreover, the theory at sub-ionization intensities 
can be done essentially exactly. 

The technical challenges of an ion beam experiment~\cite{wang:043411} 
are obvious reasons that the 
H$_2^+$ experiments have not yet been done. A more fundamental problem, however, lies
in the fact that H$_2^+$ typically comes in a broad rovibrational distribution in such
experiments~\cite{wang:043411}. 
Unfortunately, dissociation of H$_2^+$ from different initial $v$
by a linearly polarized laser pulse 
gives fragments with similar energies.
Since the asymmetry produced by each $v$ is slightly different, the incoherent averaging 
over initial $v$ required for an H$_2^+$ beam tends to wash out
the overall asymmetry~\cite{Roudnev:PRA:2007:HDp:CEP}. Moreover, one-photon dissociation of higher 
$v$ dominates the total signal. And, since one-photon dissociation produces a single nuclear parity, its 
momentum distribution is symmetric, masking the desired
asymmetry.

In this Letter, we present a scheme to greatly enhance CEP effects.
This enhancement is largely achieved by depleting the undesired higher-$v$ states with 
a long, weak pump pulse. Subsequent dissociation of this prepared state by a few-cycle 
probe pulse gives a momentum 
distribution with an order of magnitude enhanced asymmetry compared to that of an initial incoherent 
Franck-Condon distribution. In fact, our scheme gives larger asymmetries --- at longer pulse lengths ---
than have been observed so far in H$_2$ 
experiments~\cite{Kling:Science:2006:ElectronLocalization,kremer:213003,PhysRevLett.105.223001,Kling.MolPhys.2008}. 
We also propose ways to separate the pump and probe signals. 
To support our claims, we present theoretical CEP-dependent $p$+H momentum distributions in
addition to the up-down asymmetry. Calculating such a differential observable --- along with averaging over the
intensity distribution of the laser focus --- permits us to quantitatively predict the experimental 
outcome and to provide deeper physical insight.
  
To obtain the momentum distribution, it is necessary to account for the nuclear rotation. We thus solve
the time-dependent Schr\"{o}dinger equation in the Born-Oppenheimer
representation, including nuclear rotation, nuclear vibration,
and electronic excitation, but neglecting ionization as well as the Coriolis
and all nonadiabatic couplings.
The nuclear rotation is included as an expansion of the wave function over
the total orbital angular momentum ($J$) basis (see \cite{anis:033416} for details).

To prepare the system, we use a 785~nm, 45~fs long pump pulse with an intensity   
of $10^{13}$~W/cm$^2$. This relatively long, weak pump pulse depletes the
higher $v$ states, eliminating their spatially symmetric dissociation signal.  
\begin{figure}[t]
\begin{center}
\includegraphics[clip=true,width=2.5in]{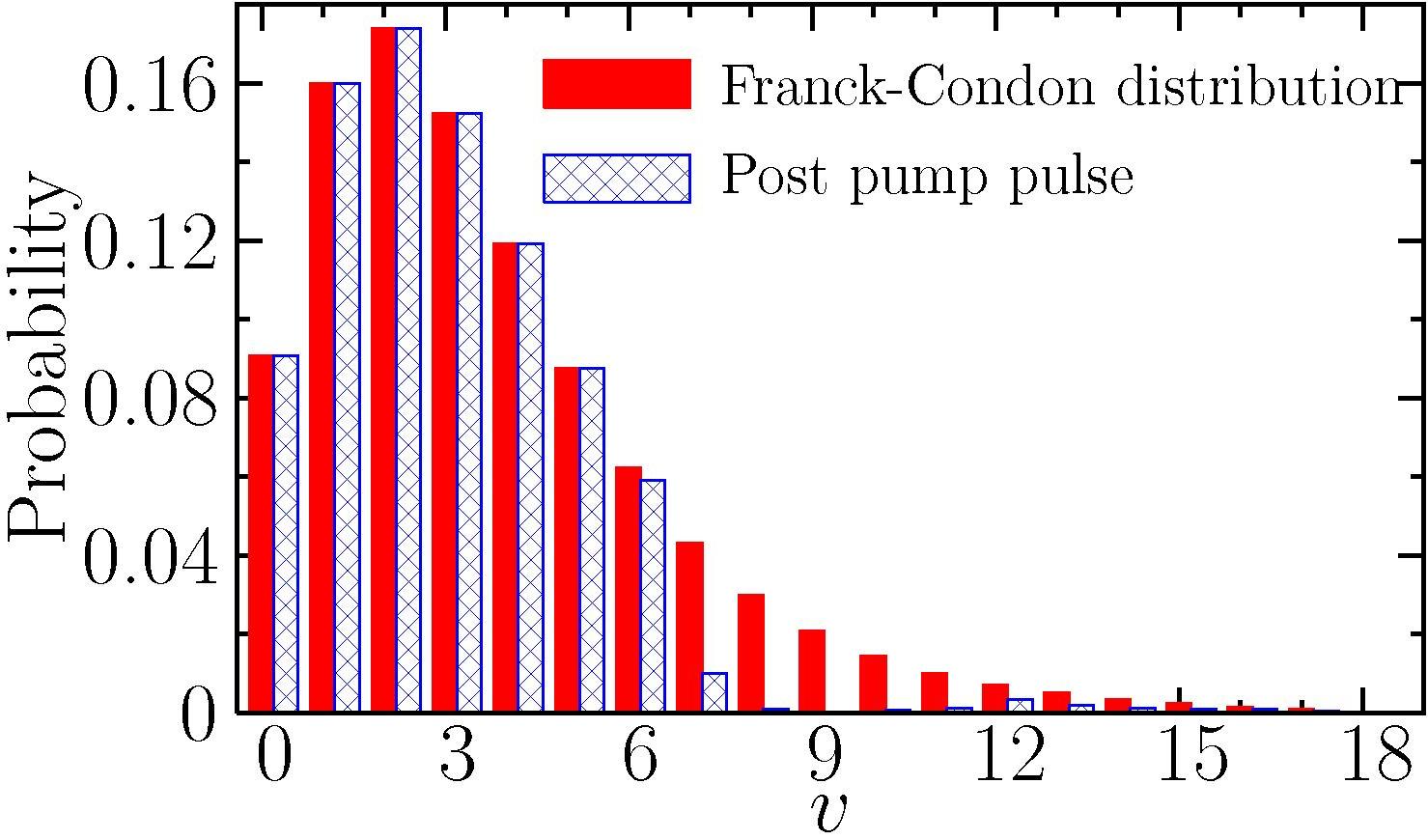}
\caption{Vibrational state distribution before and after the  45~fs, $10^{13}$ W/cm$^2$ pump pulse.}
\label{DPplot}
\end{center}
\end{figure}
Figure~\ref{DPplot} shows the $v$-distribution before and after this pump pulse.
In an incoherent Franck-Condon $v$-distribution, 9.58\% of the population 
lies in $v\geq8$. Since 90\% of this population dissociates in the pump pulse, 
$v\geq8$ becomes only 1.36\% of the total remaining bound population, 
ensuring that their contribution to the dissociation signal by any subsequent 
probe pulse will be negligible.
Consequently, we performed probe pulse calculations only including $v=$0--7, after
verifying for a representative case that $v\geq8$ affected
the asymmetry by much less than 1\%. 

To quantify the enhancement, we compare the results from
an initial incoherent Franck-Condon distribution interacting with {\em only} the probe 
pulse (``probe-only'') to
the signal from the probe part of our proposed pump-probe scheme (``pump-probe'').
We used a 7~fs, 785~nm probe pulse in both cases. 
In the pump-probe scheme, all calculations were performed at a fixed time 
delay of 267~fs unless stated otherwise. 
Since ionization is neglected, 
we limit the peak intensity to no more than $1.2\times10^{14}$~W/cm$^2$.
For the peak intensities above $10^{13}$~W/cm$^2$ required for intensity averaging, our calculations
included $p$+H($2l$) manifold in addition to $1s\sigma_{g}$ and $2p\sigma_{u}$ channels. 
The total population of the $p$+H($2l$) states was less than 5\% even for the highest intensity. 
Consequently, we present momentum distributions 
based on just the $1s\sigma_{g}$ and $2p\sigma_{u}$ channels.  

The fundamental physical observable we focus on is the $p$+H relative momentum distribution 
$\rho(\bf K)$, which is the most differential observable 
in recent experiments involving H$_2^+$ dissociation~\cite{McKenna:PRL:2008:H2:10fs:ATD,wang:043411,BenItzik:PRL:2005:H2}. 
To calculate $\rho(\bf K)$, we project the final wave function onto scattering
states that behave as $\exp(i{\bf K}\cdot{\bf R}) \phi_{1sA}$ asymptotically,
where ${\bf R}$ points from proton $A$ to proton $B$ and $\phi_{1sA}$ is the 
hydrogen ground state wave function centered on proton $A$.  The momentum $\bf K$ thus points
from H to $p$.  This scattering state, with the nuclear spin included, is then 
symmetrized to account for the identical nuclei~\cite{Green.PRL.21.1732,singer:6060,AnisThesis}.  
Finally, the momentum distribution [or its energy-normalized equivalent $\rho(E,\hat{K})$
with $E=K^{2}/2\mu$, $\mu$ the nuclear reduced mass, and $\hat{K}=(\theta_{K},\varphi_{K})$ 
the direction of $\bf K$ with respect to the polarization direction] is 
\begin{align}
\rho({\bf K}) 
=&\frac{1}{\mu\sqrt{2\mu E}}\rho(E,\hat{K})
\label{EqMD}
\\
=&\frac{1}{\mu\sqrt{2\mu E}}\biggl|\sum_{J\,{\rm even}}C_{Jg}Y_{JM}(\hat{K}) 
\!+\!\!\! 
\sum_{J\,{\rm odd}}C_{Ju}Y_{JM}(\hat{K})\biggr|^{2}
\nonumber
\end{align}
with
\begin{align} 
C_{Jp}= (-i)^{J}e^{-i\delta_{Jp}}\langle EJp|F_{Jp}(t_{f})\rangle,\qquad p=g,u.
\label{EqCoef}
\end{align}
Here, $|F_{Jp}(t_{f})\rangle$ are the $1s\sigma_{g}$ and $2p\sigma_{u}$
nuclear radial wave functions at the final time $t_{f}$, while
$|EJp\rangle$ and $\delta_{Jp}$ are the corresponding energy-normalized scattering states and
phase shifts, respectively.  

Equation~(\ref{EqMD}) shows that although a linear combination
of $1s\sigma_{g}$ and $2p\sigma_{u}$ is necessary to localize the electron
as an atomic rather than a molecular state, the spatial asymmetry of $p$+H
is due to the interference of even and odd nuclear parity states.  This
distinction is brought into sharp relief when nuclear rotation is included
in the calculation since using simply $1s\sigma_{g}$$\pm$$2p\sigma_{u}$ nuclear wave function---
as is done in calculations without rotation --- would produce two distinct $p$+H
momentum distributions, where clearly only one can be measured.  It is
the symmetrization requirement that dictates the proper coherent combination to use.
This issue is not new, however, and always arises for identical particle
scattering where it is known that the primary differences occur for $\theta_K$$\approx$$\pi/2$.
Since intense-field dissociation of H$_2^+$ produces very few fragments
at this $\theta_K$, the consequences of analyzing incorrectly are less pronounced.
For more complicated systems, however, this need no longer be true.

\begin{figure}[!tb]
\begin{center}
\includegraphics[clip=true,height=30mm]{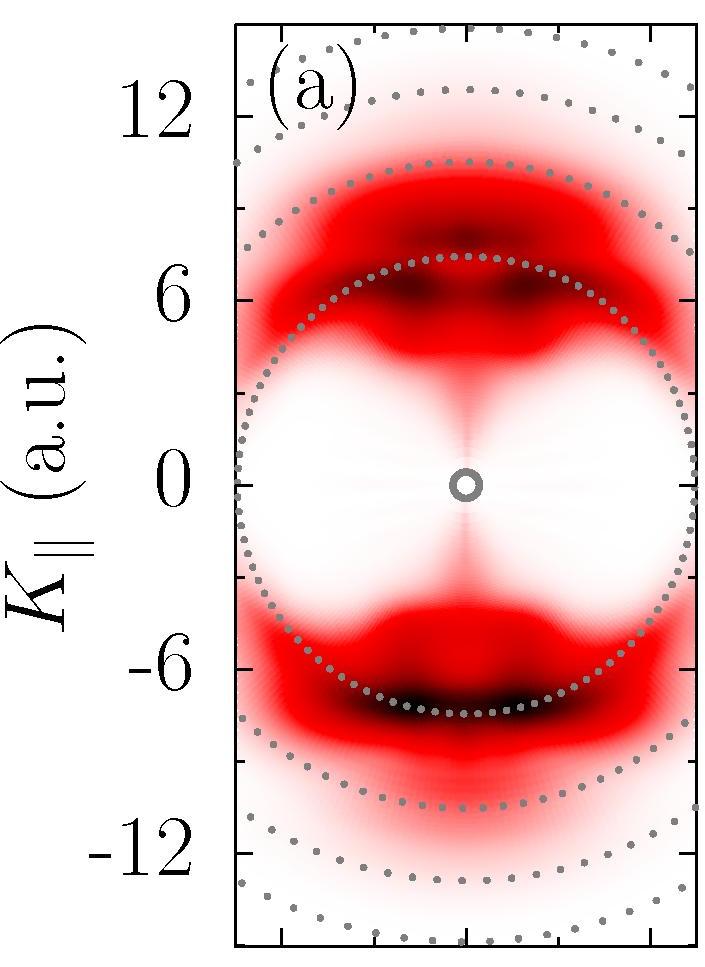}
\includegraphics[clip=true,height=30mm]{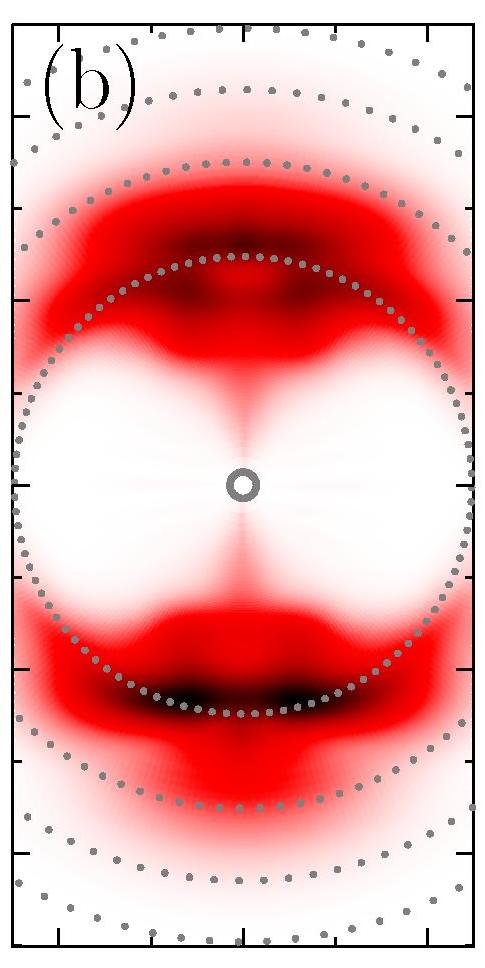}
\includegraphics[clip=true,height=30mm]{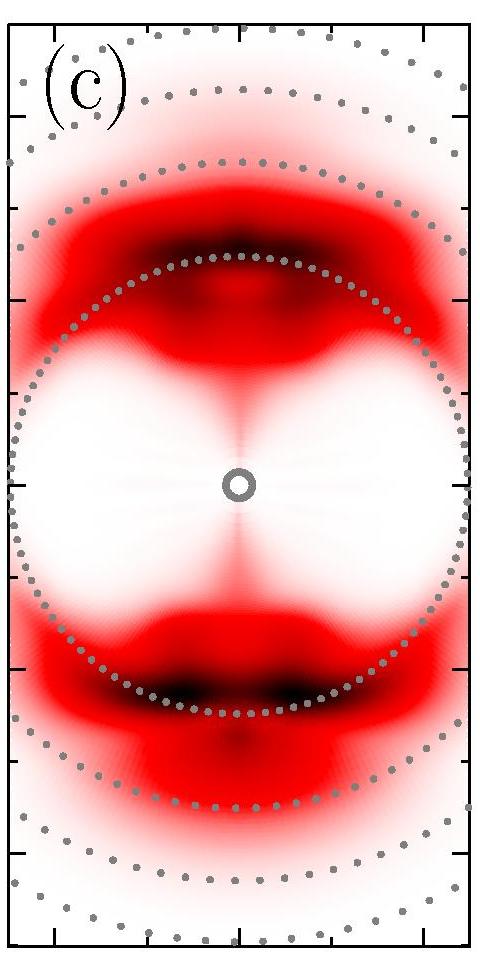}
\includegraphics[clip=true,height=30mm]{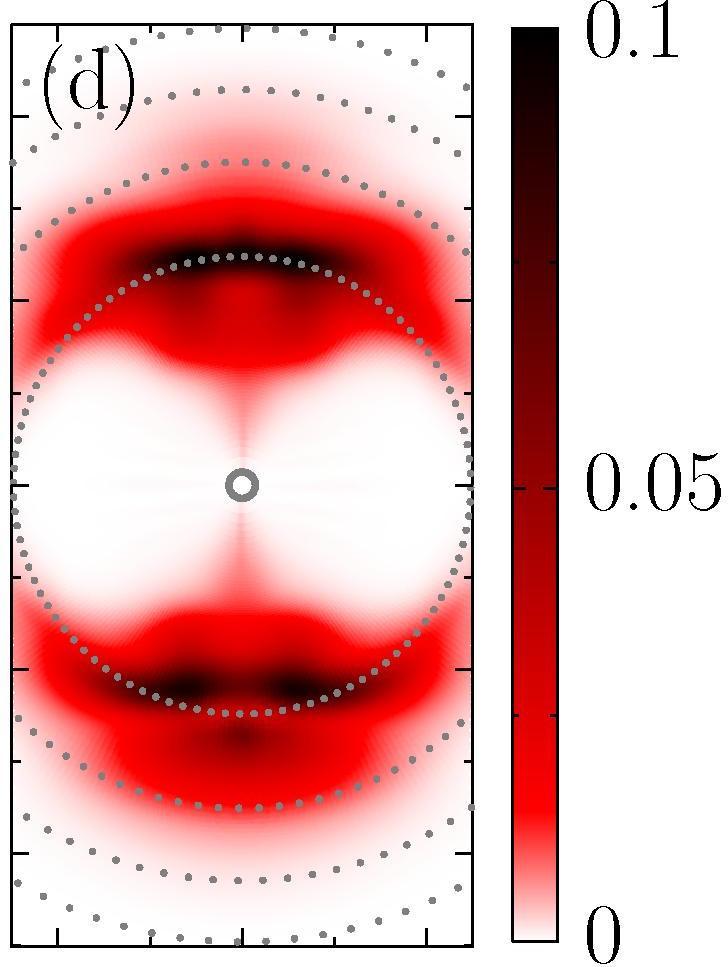}
\\
\includegraphics[clip=true,height=36.508mm]{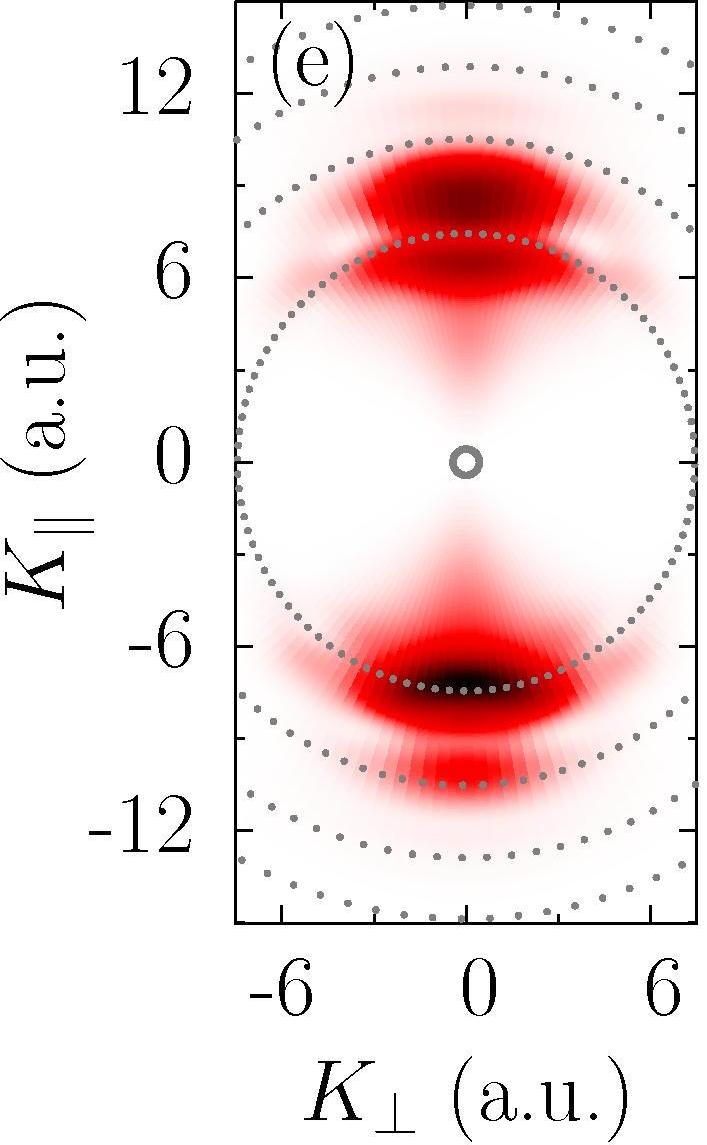}
\includegraphics[clip=true,height=36.508mm]{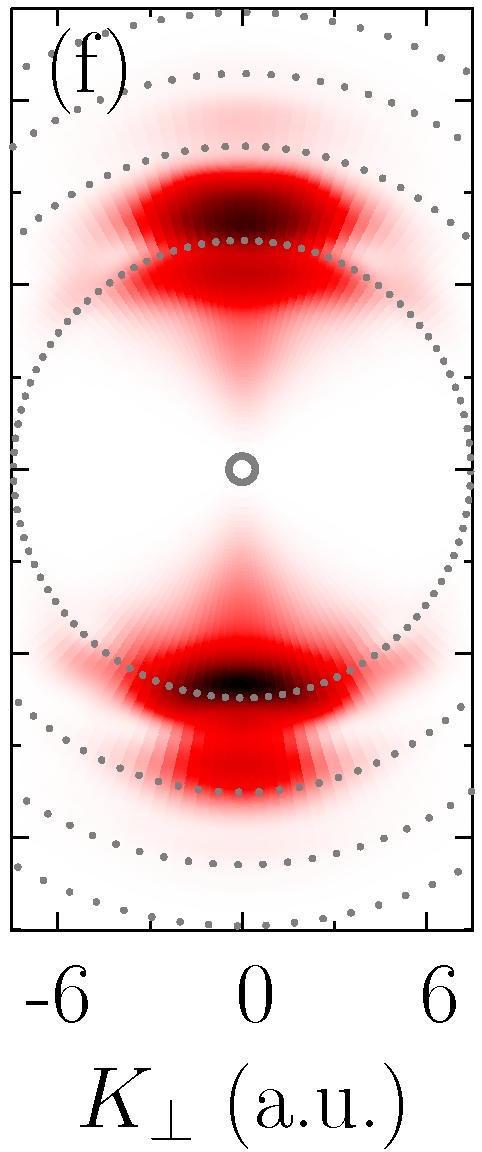}
\includegraphics[clip=true,height=36.508mm]{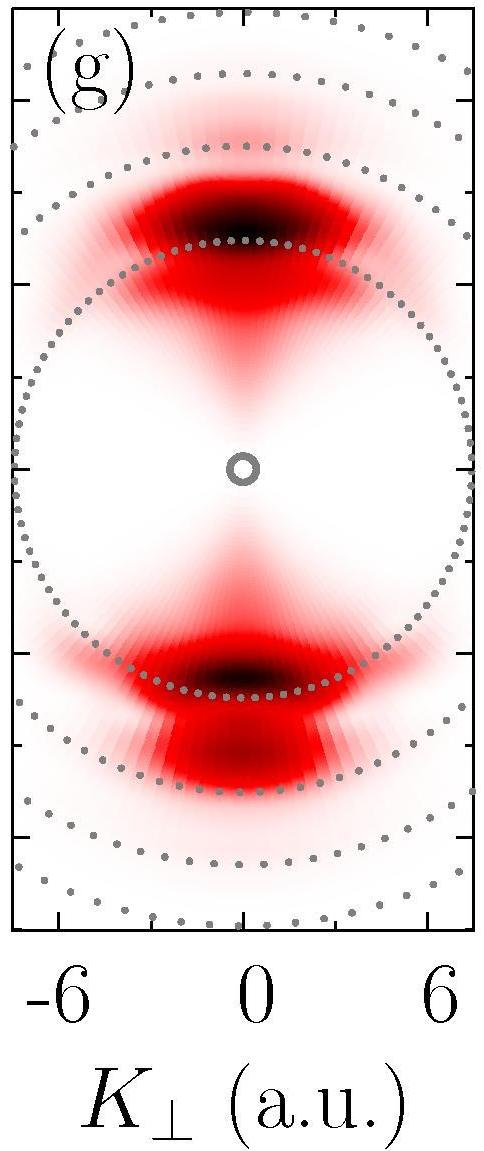}
\includegraphics[clip=true,height=36.508mm]{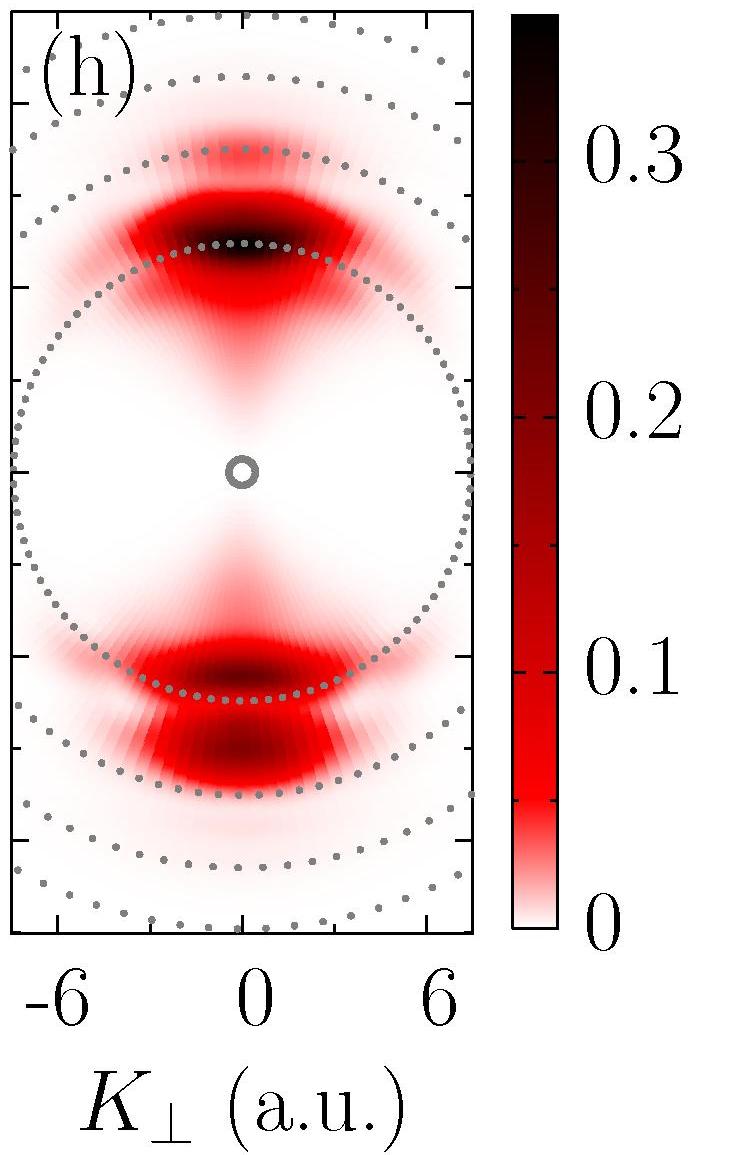}
\end{center}
\caption{Franck-Condon-averaged $K^{2}\rho({\bf K})$ (integrated over $\varphi_{K}$) 
for probe-only [$\rho(\bf K)$ is reflected to $-K_{\perp}$ for clarity] for (a) $\varphi =0$,
(b) $\varphi=\pi/4$, (c) $\varphi=\pi/2$, and (d) $\varphi=3\pi/4$ (
Gray dotted lines mark $K$ = 7.5, 10, 12.5, and 15~ a.u.). 
(e)--(h) are same as (a)--(d) but for pump-probe. All cases used a 
7~fs, $10^{14}$~W/cm$^2$ probe pulse.}
\label{MomD}
\end{figure}

The Franck-Condon-averaged momentum distributions for several CEPs 
are shown in Figs.~\ref{MomD}(a)--(d) for the probe-only case and in 
Figs.~\ref{MomD}(e)--(h) for the pump-probe case. 
The momentum distributions in all cases exhibit preferential alignment along the laser 
polarization. Moreover, since the energy distributions
\begin{align} 
\rho(E)=\int\rho(E,\hat{K})d\Omega_{K}
\label{EqKER}
\end{align}
for the $1s\sigma_{g}$ and $2p\sigma_{u}$ channels of individual vibrational states 
overlap roughly in the range 0.5--1.5~eV, we expect spatial asymmetries to appear 
roughly for 6$\leq$$K$$\leq$10~a.u.  The results shown in Fig.~\ref{MomD}
for both experimental scenarios are consistent with this expectation.
The momentum distribution for $\varphi=\varphi+\pi$ 
is the mirror image of the momentum distribution for $\varphi$, as guaranteed by the fact that 
$\cos(\omega t+\pi)=-\cos\omega t$ in Eq.~(\ref{pulse}).

While the two experimental scenarios clearly show qualitative differences, 
the strikingly different distributions make it difficult to judge which
produces the larger asymmetry.  We thus turn to the
quantitative measure of the asymmetry used in previous studies \cite{Kling:Science:2006:ElectronLocalization,Hua:JPB:2009,Kling.MolPhys.2008}: the
normalized asymmetry parameter ${\cal A}(E,\varphi)$, 
\begin{align}
{\cal A}(E,\varphi) = \rho(E)^{-1}\left[\rho(E)_{\rm Up}
-\rho(E)_{\rm Down}\right].
\label{AsymEq}
\end{align}
For simplicity, we integrate over the whole upper and lower hemispheres in
the ``Up'' and ``Down'' distributions, respectively, although a narrow 
angular cut along the laser polarization direction might be chosen
to enhance $\cal A$ as in some experimental
studies \cite{Kling:Science:2006:ElectronLocalization,Kling.MolPhys.2008}. 
Figure~\ref{MomD} shows why such cuts are effective since the strongest 
CEP dependence lies at small $\theta_{K}$.  Although
the total energy spectrum $\rho(E)$ in principle also depends 
on CEP~\cite{Hua:JPB:2009}, we found negligible  CEP-dependence in the Franck-Condon averaged $\rho(E)$
and thus expect essentially no contribution to the CEP dependence from the denominator of $\cal A$.

\begin{figure}[!tb]
\begin{center}
\includegraphics[clip=true,height=1.in]{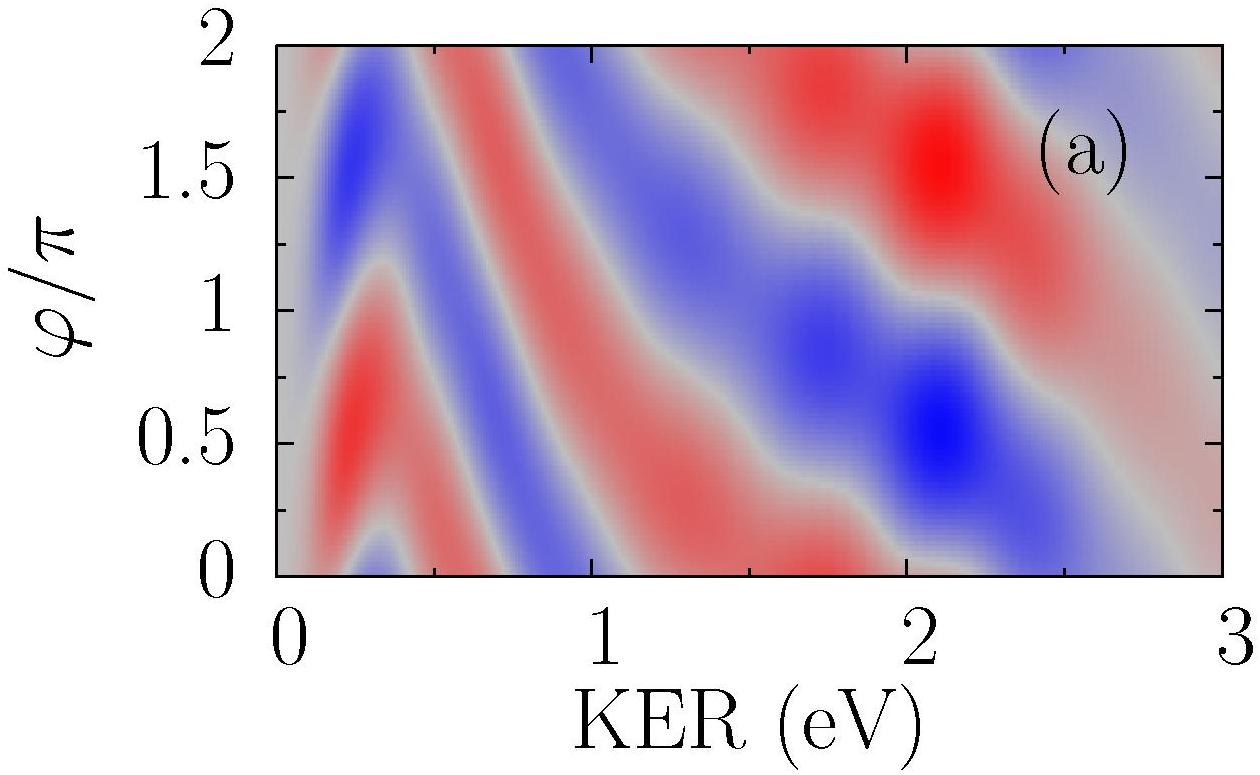}
\includegraphics[clip=true,height=1.in]{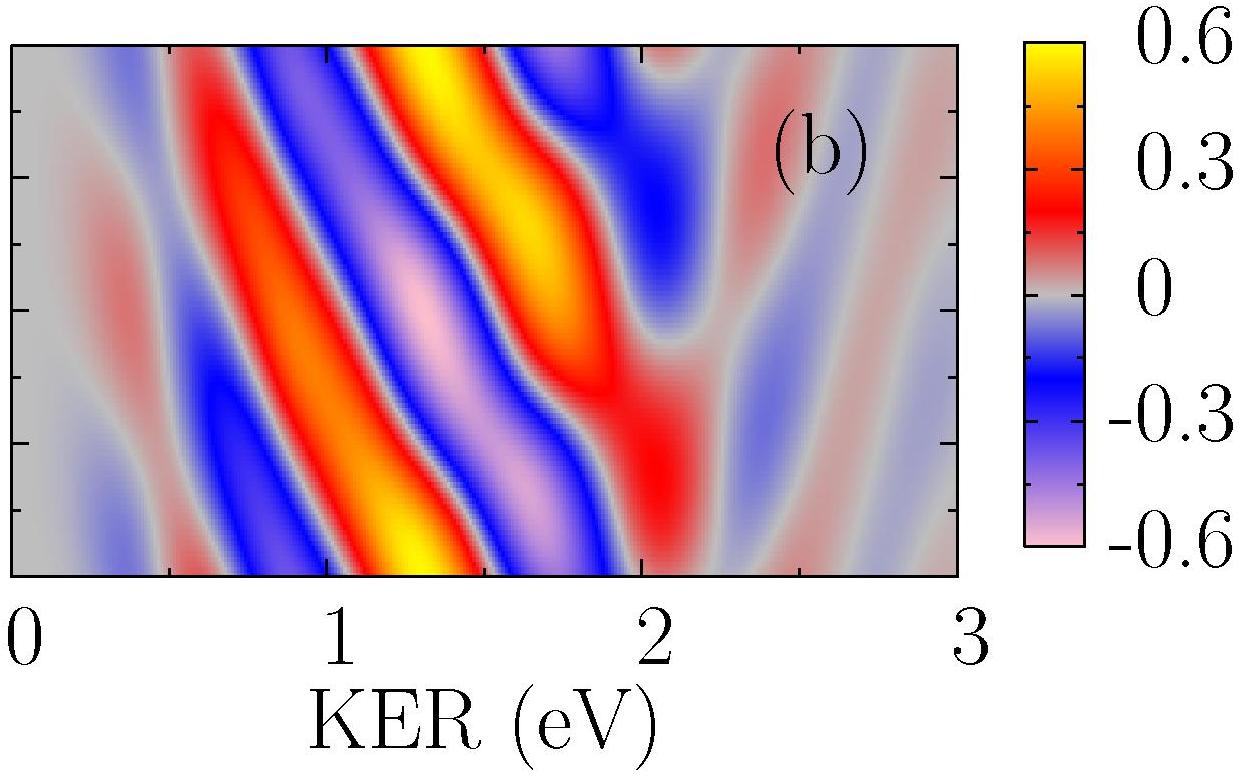}
\end{center}
\caption{(a) Asymmetry defined in Eq.~(\ref{AsymEq})  
for the probe-only case and (b) for the pump-probe case for a $\tau_{\rm FWHM} = 7$~fs and $I=10^{14}$~W/cm$^2$ pulse.}
\label{Asymm}
\end{figure}

Figures~\ref{Asymm}(a) and \ref{Asymm}(b) show ${\cal A}(E,\varphi)$ for the 
proble-only and the pump-probe, respectively. 
For the probe-only in Fig.~\ref{Asymm}(a), we can already see reasonable asymmetry 
in the range 0.2--2.5~eV where it oscillates between $-0.12$ and $0.12$.
Comparing Fig.~\ref{Asymm}(a)
and Fig.~\ref{Asymm}(b), however, we find a five-fold enhancement of $|{\cal A}(E,\varphi)|$ 
in the pump-probe case for this intensity. 

The most crucial factor determining whether an intensity-dependent effect is experimentally observable 
is whether it survives intensity (or focal volume) averaging. We thus  
intensity-averaged our results for a 7~fs laser pulse with $\varphi=0$ and a peak intensity of
$1.2\times10^{14}$~W/cm$^2$. 
We used the two-dimensional geometry
of Ref.~\cite{McKenna:PRL:2008:H2:10fs:ATD} to perform the intensity-averaging
following the procedure described in \cite{Roudnev:PRA:2007:HDp:CEP}.
Figures~\ref{MomD}(a), \ref{MomD}(e) and \ref{Asymm}, 
show a clear up-down asymmetry for $\varphi =0$ in both probe-only and pump-probe 
cases, and we will check if it survives intensity averaging. 
\begin{figure}[!b]
\begin{center}
\includegraphics[width=3.3in,clip=true]{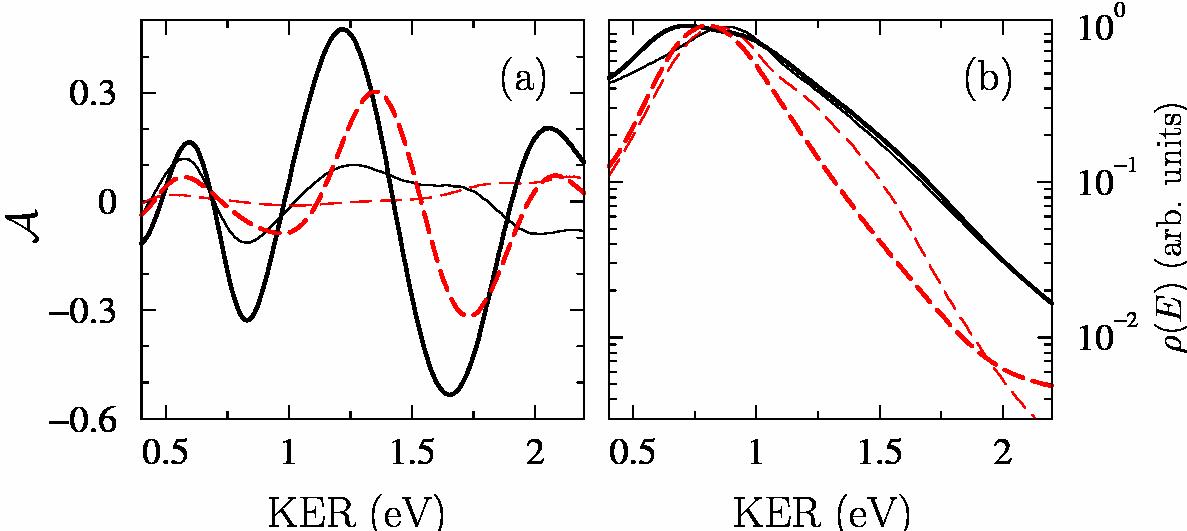}
\caption{(a) Asymmetry from
the intensity-averaged $\rho(E,\hat{K})$ for the
probe-only (thin dashed lines) and the pump-probe (thick dashed lines) cases as well as for a single 7~fs probe pulse 
with a peak intensity of
$1.2\times10^{14}$~W/cm$^2$ (thin and thick solid lines, respectively). 
(b) KER distributions for the cases shown in (a)
normalized to the same peak value.}
\label{IntAvg}
\end{center}
\end{figure}
Figure~\ref{IntAvg}(a) shows ${\cal A}(E,0)$ before and 
after the intensity averaging for both cases. 
The corresponding total KER distributions are plotted in
Fig.~\ref{IntAvg}(b) to show that the small $\rho(E)$ is the reason for 
the large ${\cal A}(E,0)$ at higher energies. 
In the pump-probe case, the intensity averaging has only been performed over the probe-pulse
intensity distribution assuming that the weak pump intensity can be made 
uniform across the probe focal volume.

For the probe-only case, intensity averaging 
reduces $\cal A$ by more than a factor of three over the entire energy range 
shown in Fig.~\ref{IntAvg}(a) and makes it 17 times
smaller for 0.5 to 1.0 eV, where $\rho(E)$ is large and the single-intensity $\cal A$
is largest. This significant reduction in $\cal A$ is due to the  
fact that one-photon dissociation --- which shows no asymmetry ---
can occur for $v\geq7$ at very low intensities ($\approx$10$^{10}$~W/cm$^2$).
These symmetric contributions are thus amplified by the intensity averaging and swamp any asymmetry
because essentially all $v$ contribute to these KER.
Figure~\ref{IntAvg}(a) thus shows that intensity averaging makes it very challenging to
measure CEP-effects for a single 7~fs or longer pulse in an experiment.

For the pump-probe case, $\cal A$ is also reduced from the single intensity 
value --- but to a much lesser extent than in the probe-only case.
In fact, Fig.~\ref{IntAvg}(a) shows that even after intensity averaging
$\cal A$ is an order of magnitude larger using the pump-probe scheme compared 
to the probe-only results.  Moreover, we have found that the 
pump-probe scheme produces a CEP-dependent asymmetry after intensity averaging
even for 10~fs pulses. 
These pulses are much longer than the 6~fs pulses that have been used to date to 
observe CEP effects \cite{Kling:Science:2006:ElectronLocalization,kremer:213003}. 

Besides depleting the high-lying vibrational states, the pump also 
impulsively aligns the molecule~\cite{PhysRevA.2011,PhysRevA.77.053407,RevModPhys.75.543}). 
To investigate the sensitivity of $\cal A$ to the alignment, 
we calculated the asymmetry for three different pump-probe time delays with 
aligned $\langle\cos^{2}\theta\rangle$=0.56, 
anti-aligned $\langle\cos^{2}\theta\rangle$=0.22,
and dephased $\langle\cos^{2}\theta\rangle$=0.40 angular distributions 
($\langle\cos^{2}\theta\rangle$=1/3 for an isotropic distribution). 
We found that the maximum $\cal A$ was largest for the aligned distribution, followed 
by the anti-aligned, with the dephased smallest.  The enhancement of the aligned $\cal A$
over the dephased was roughly 30\%.
For this reason we have shown here calculations for the 
267~fs delay corresponding to the aligned distribution.
This exercise also served to establish that the major source
of the ten-fold CEP-dependent asymmetry enhancement is the depletion of
the higher-$v$ states. 

Another concern for experimentally observing the predicted enhancement is the fact
that in our pump-probe scheme the pump pulse already produces fragments.  So,
separating the probe signal from the pump signal is crucial.
Although the dissociating fragments from both pulses
overlap in momentum, we expect the asymmetry will still be large in a combined
pump-probe signal for two reasons. 
First, the momentum distribution from the long pump pulse
exhibits narrow peaks corresponding to higher vibrational states. 
Therefore, in the combined pump-probe momentum
distribution, the symmetric structure would be very localized in KER, giving small overlap with the 
broad asymmetric signal resulting in larger asymmetry than the probe-only case.
Second, we found that preparing the initial state greatly increased the 
dissociation probability of the lower vibrational states for aligned (2.33 fold) and dephased (1.74 fold)
pump-probe cases compared to the probe-only, thereby 
enhancing the ratio of the asymmetric signal to the symmetric signal.

The contrast between pump and probe signals can be further improved over the present 
case using pump pulses longer than 45~fs, thus
increasing the depletion of the higher vibrational states and making the pump signal
even more structured. A longer pulse will give more alignment, which might also enhance asymmetry.
Additionally, instead of using the whole upper and lower hemispheres to define $\cal A$, an angular cut 
can be used to isolate the aligned asymmetric distribution. 

As the dissociating fragments primarily lie along the laser polarization, 
it might be better to use orthogonal laser polarization
directions for the pump and probe pulses to separate their signals~\cite{PhysRevLett.105.223001}. 
For this, one might want to use the time-delay when the molecules are antialigned relative to
the pump polarization to improve the signal.  A circularly polarized pump pulse could
also be used.
Since depletion is the major reason for enhanced CEP effects, we believe
the effect will survive using different laser polarizations.
An intensity differencing scheme might also be useful to enhance
asymmetry~\cite{Wang:OL:2005:IDS}.

In this Letter, we have presented a prescription for substantially enhancing CEP effects
on the spatial asymmetry of intense-field-induced fragmentation. We have illustrated our proposal with 
essentially exact calculations for the benchmark system H$_2^+$, including the important 
averaging implicit in experiments and found a ten-fold increase in the asymmetry.
In addition, we have suggested several other steps that could further increase the asymmetry.

While even greater enhancement could be realized by 
preparing the system in a single initial $v$,
our scheme provides a more easily followed experimental avenue yielding 
a narrow vibrational state distribution.
We believe that our scheme is equally applicable to neutral molecules with studies focussed on dissociative 
ionization channels, where the initial intense long pump pulses can serve to ionize, dissociate undesired 
vibrational states, and align the molecular ions. 

We gratefully acknowledge many useful discussions with I. Ben-Itzhak and J. McKenna regarding
experimental limitations.
The work was supported by the Chemical Sciences, Geo-Sciences, and Biosciences
Division, Office of Basic Energy Sciences, Office of Science, U.S. Department
of Energy.

\bibliography{CEPeffects.New.3N}

\begin{thebibliography}{37}
\bibitem[{Par()}]{Param}
\bibinfo{note}{The laser electric field ${\cal E}_{0}$ in atomic units is
  related to the peak intensity $I$ by ${\cal
  E}_{0}=\sqrt{I/(3.5\times10^{16}{\rm W/cm}^{2})}$, $\tau$ is related to the
  full width of the intensity at half maximum $\tau_{\rm FWHM}$ by
  $\tau=\tau_{\rm FWHM}/\sqrt{2\ln2}$, and $\omega$ is the frequency in atomic
  units.}
\end{thebibliography}
    
\end{document}